\documentclass[aip,jap,reprint]{revtex4-1}
\usepackage{graphicx}
\DeclareGraphicsExtensions{.pdf,.jpg,.png,.eps}
\usepackage{amsfonts}
\usepackage{amsmath}
\usepackage{color}
\begin{document}
\title{Glassy Dielectric Response in Tb$_{2}$NiMnO$_{6}$ Double Perovskite with Similarities to a Griffiths Phase}
\author{Hariharan N.$^*$}
\email{hariharan.nhalil@gmail.com}
\affiliation{Department of Physics, Indian Institute of Science, Bangalore 560012, India}
\author{Harikrishnan S. Nair$^*$}
\email[] {h.nair@fz-juelich.de, krishnair1@gmail.com\\ $^*$ authors contributed equally to the work}
\affiliation{J\"{u}lich Center for Neutron Sciences and Peter Gr\"{u}nberg Institute, JARA-FIT, Forschungszentrum J\"{u}lich GmbH, 52425 J\"{u}lich, Germany}
\affiliation{Department of Physics, Indian Institute of Science, Bangalore 560012, India}
\author{H. L. Bhat}
\affiliation{Department of Physics, Indian Institute of Science, Bangalore 560012, India}
\affiliation{Center for Soft Mater Research, Jalahalli, Bangalore 560013, India}
\author{Suja Elizabeth}
\affiliation{Department of Physics, Indian Institute of Science, Bangalore 560012, India}
\date{\today}
\begin{abstract}
Frequency-dependent and temperature-dependent dielectric measurements are performed on double perovskite Tb$_2$NiMnO$_6$. The real ($\epsilon_1$) and imaginary ($\epsilon_2$) parts of dielectric permittivity show three plateaus suggesting dielectric relaxation originating from bulk, grain boundaries and the sample-electrode interfaces respectively. The temperature and frequency variation of $\epsilon_1$ and $\epsilon_2$ are successfully simulated by a $RC$ circuit model. The complex plane of impedance, $Z'$-$Z"$, is simulated using a series network with a resistor $R$ and a constant phase element. Through the analysis of frequency-dependent dielectric constant using modified-Debye model, different relaxation regimes are identified. Temperature dependence of dc conductivity also presents a clear change in slope at, $T^*$. Interestingly, $T^*$ compares with the temperature at which an anomaly occurs in the phonon modes and the Griffiths temperature  for this compound. The components $R$ and $C$ corresponding to the bulk and the parameter $\alpha$ from modified-Debye fit tend support to this hypothesis. Though these results cannot be interpreted as magnetoelectric coupling, the relationship between lattice and magnetism is marked.
\end{abstract}
%
\maketitle
\section{Introduction}
Double perovskite compounds $R_2BB'$O$_6$ ($R$ = rare earth; $B,B'$ = transition metal) display a variety of interesting properties such as ferromagnetism
\cite{Dass2003}, magnetocapacitance/ magnetoresistance \cite{Rogado2005} and field-induced changes of dielectric constant \cite{Singh2007} all  which make them potential candidates for spintronics applications. Theoretical predictions \cite{Kumar2010}
and experimental observation \cite{Yanez-Vilar2011} of multiferroicity have been reported for double perovskites. However, most investigations of this class of compounds were focused on La-based compositions, for example, La$_2$NiMnO$_6$ which has a high ferromagnetic transition temperature of 280~K \cite{Padhan2008,lin_ssc_149_784_2009}. Ceramics of La$_2$NiMnO$_6$ are reported to show relaxor-like dielectric response which is attributed to Ni$^{2+}$ -- Mn$^{4+}$ charge ordering \cite{lin_ssc_149_784_2009}. Epitaxial thin films of La$_2$NiMnO$_6$ are known for dielectric relaxation and magnetodielectric effect\cite{Padhan2008}. In the present paper, we report the results of impedance spectroscopy of Tb$_2$NiMnO$_6$. The motivation for the this work stems from our previous studiy that showed a clear correlation between lattice anomaly observed in FWHM of phonon mode and  Griffiths temperature observed through magnetization \cite{Nair2011}. Here we address  the dielectric response of this system, its interpretation and appraisal  based on the previous knowledge. Experimentally observed dielectric data is faithfully reproduced using resistor network models which help to extract  intrinsic contributions. A characteristic temperature is identified in the ensuing  analyses which is compared with the magnetic and Raman data already reported on this material. It is found that the magnetic Griffiths temperature is reflected in the dielectric data also through this characteristic temperature.
\\
\begin{figure}[!t]
\centering
\includegraphics[scale=0.50]{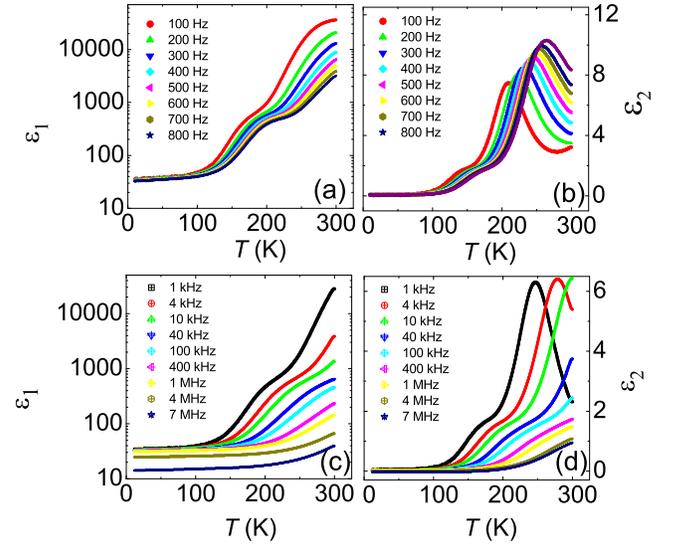}
\caption{(colour online) The real part of dielectric permittivity, $\epsilon_1$, and the dissipation factor or loss, $\epsilon_2$, as a function of temperature, measured with different applied frequencies in the range 100~Hz -- 800~Hz (a, b) and 1~kHz -- 7~MHz (c, d). In the low frequency region, three different plateaus are observed which can arise from intrinsic, grain boundary and sample-electrode interfaces respectively.}
\label{fig_epsilon}
\end{figure}
\section{Experimental Details}
Details of  synthesis, structure, magnetism and Raman studies of Tb$_2$NiMnO$_6$ were reported earlier \cite{Nair2011}. In order to perform dielectric measurements, pellets of approximate thickness 0.8~mm and area 6.8~mm$^2$ were prepared using poly-vinyl alcohol as a binder. Density of the pellet is measured to be greater than 95 $\%$ of the theoretical density. Temperature dependent dielectric constant was measured using a Janis cryostat in the frequency range 1~kHz to 10~MHz using a 4294A precision impedance analyser with an applied ac voltage of 800~mV. Dielectric experiments on these samples were repeated using several electrodes. Initially, silver paste was applied on both sides of the pellet and was baked at 250$^{\circ}$C for 3-4~hours before measurement. Afterwards, the measurements were repeated using silver and gold plated electrodes. The data obtained with all the three types of electrodes were consistent.
\\
\section{Results and Discussion}
\begin{figure}[!b]
\centering
\includegraphics[scale=0.35]{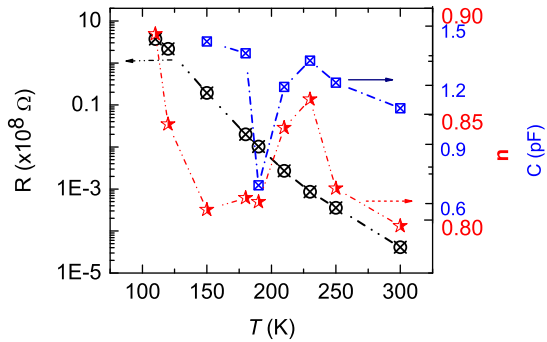}
\caption{(colour online) The estimated values of $R$, $C$ and $n$ by separating the intrinsic dielectric relaxation after analysis using the resistor-capacitor models. A broad anomaly is present centered around 200~K (the error bars were comparable to the size of the data points).}
\label{fig_R}
\end{figure}

\begin{figure}[!t]
\centering
\includegraphics[scale=0.50]{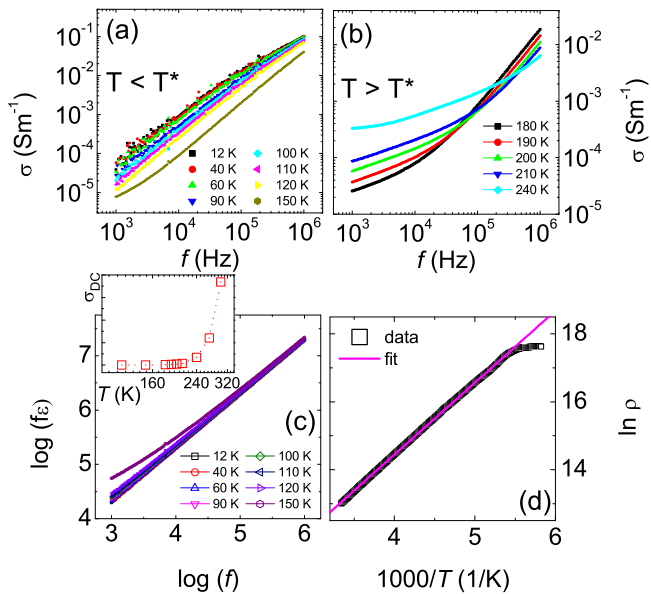}
\caption{(colour online) (a) Shows the frequency dependence of ac conductivity below $T^* \sim$ 200 K. (b) Frequency dependence of ac conductivity at  temperatures above $T^*$. (c) Plot of log ($f\epsilon_1'$) vs log ($f$) for selected fixed temperatures. The inset presents temperature evolution of dc conductivity which shows a significant change in slope close to 200~K. (d) Electrical resistance of Tb$_2$NiMnO$_6$ which conforms to Arrhenius type behaviour.}
\label{fig_accond}
\end{figure}

The temperature dependence of real and imaginary parts of dielectric permittivity, $\epsilon_1(f,T)$ and $\epsilon_2(f,T)$ of Tb$_2$NiMnO$_6$ in the frequency
region, 100~Hz-800~Hz is shown in Fig~\ref{fig_epsilon} (a, b) and for the range, 1~kHz-7~MHz in Fig~\ref{fig_epsilon} (c, d). Clear frequency dispersion is observed in both plots $\epsilon_1(f,T)$ and $\epsilon_2(f,T)$ and a closer examination reveals different plateaus. 
\begin{figure}[!t]
\centering
\includegraphics[scale=0.45]{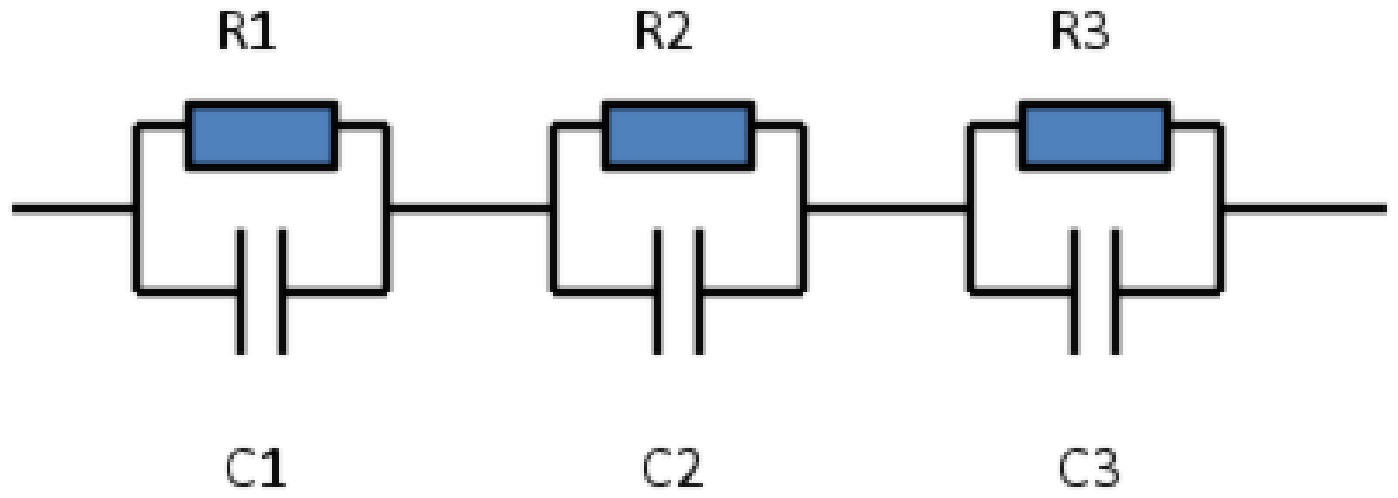}
\vspace{1cm}
\includegraphics[scale=0.17]{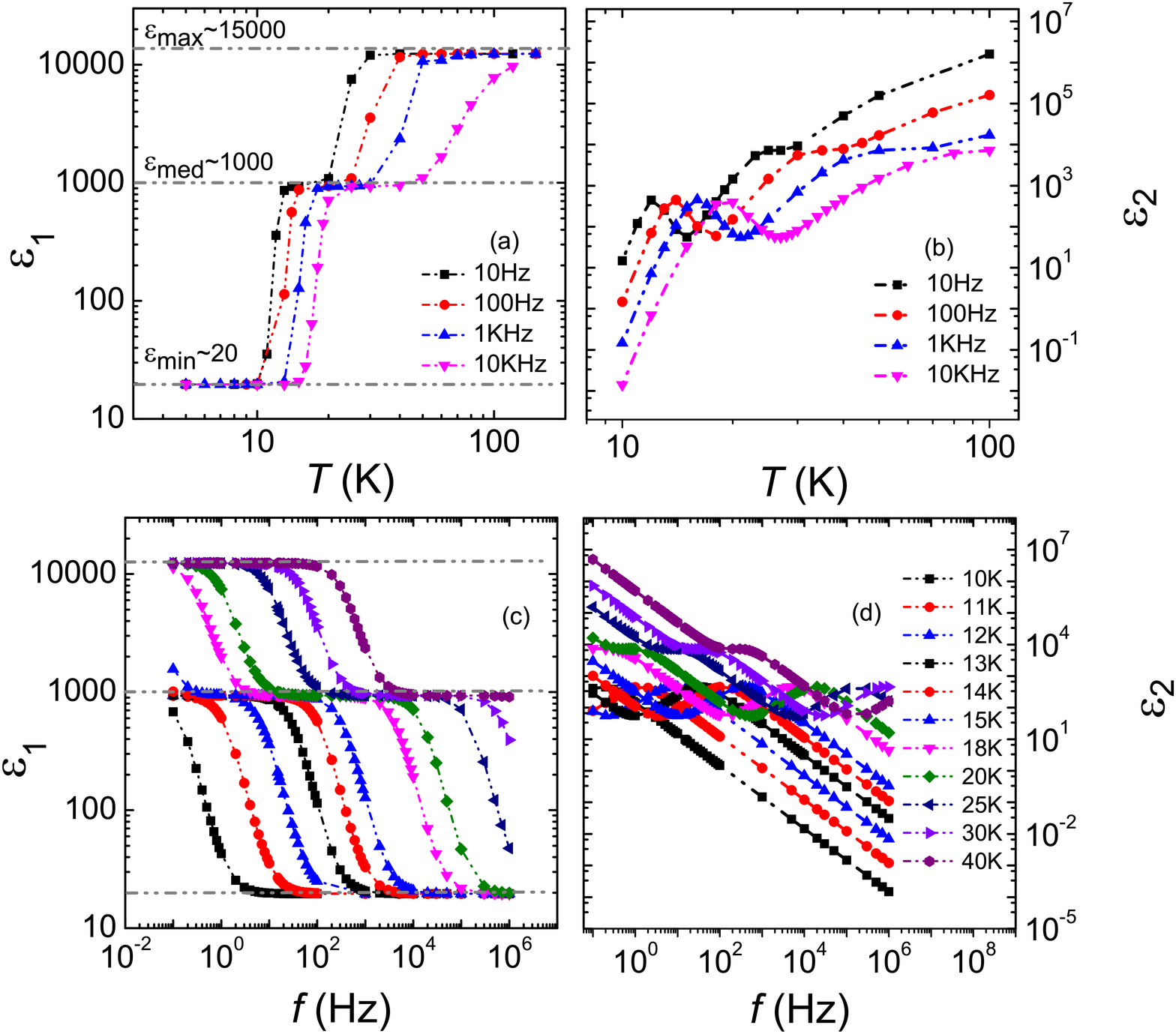}
\caption{(colour online) {\it Top:} The schematic of the $RC$ circuit used for  simulating $\epsilon_1$ and $\epsilon_2$ versus temperature plots for different frequencies. {\it Bottom:} Simulated curves of (a) $\epsilon_1$ and (b) $\epsilon_2$ versus temperature for different frequencies. As clear from (a), three plateau-regions corresponding to $\epsilon_{min}$, $\epsilon_{med}$ and $\epsilon_{max}$ are present in the experimental data also (Fig~\ref{fig_epsilon} (a, b)). Simulated curves of $\epsilon_1$ and $\epsilon_2$ as a function of frequencies for different temperatures are shown in (c) and (d).}
\label{fig_simulation}
\end{figure}
Schmidt {\it et al} \cite{Schmidt2012} have shown that  each relaxation is represented by a dielectric plateau, three of which are indeed seen  at low frequencies (Fig\ref{fig_epsilon}(a)). At high frequencies (Fig\ref{fig_epsilon}(c)) the third plateau is not observed. The low-temperature plateau originates from the intrinsic bulk contribution and the high temperature plateaus could be due to the grain boundaries and sample-electrode interface.\cite{Baron-Gonzalez2011,Schmidt2007} Each relaxation can be ideally represented by one $RC$ element\cite{Lunkenheimer2010} where $R$ and $C$ connected in parallel, and three $RC$ elements  in series (upper panel of Fig~\ref{fig_simulation}) was used to model the real and imaginary part of the permittivity versus $T$ and frequency dependence. 
\\
\indent The complex impedance is given by\cite{Macdonald-Book}
\begin{equation}
Z^*=Z'+iZ"
\end{equation} 
where $Z'$ and $Z"$ are real and imaginary part of the $Z^*$ respectively.\\
For a series of three $RC$ elements as shown in Fig~\ref{fig_simulation}, the real and imaginary part of the permitivity are calculated\cite{Schmidt2012} as ,
\begin{eqnarray}
Z^* = \frac{R_1}{1 + i\omega R_1C_1} + \frac{R_2}{1 + i\omega R_2C_2} + \frac{R_3}{1 + i\omega R_3C_3} \\
\epsilon_1 = \frac{-Z"}{\omega \epsilon_0 g \left(Z'^2 + Z"^2 \right) } \\
\epsilon_2 = \frac{Z'}{\omega \epsilon_0 g \left(Z'^2 + Z"^2 \right) } 
\end{eqnarray}
where $\omega$ is the angular frequency of the applied ac voltage and $\epsilon_0$ is the permittivity of free space and $g$ is a geometrical factor ($\propto$ area/ thickness). In equation (2), $R_1C_1$ represent the intrinsic contribution while $R_2C_2$ and $R_3C_3$ are the external contributions from the grain boundaries and the sample-electrode interface respectively. Initially all capacitances are treated as temperature independent and the resistors  obey Arrhenius type activated behavior \cite{Schmidt2012} $R_n$= $a_n$~exp(20~meV/$k_B$T). Assuming $R_3 \gg R_2 \gg R_1$, the pre-exponential constants $a_1$, $a_2$ and $a_3$ are taken as 1, 1000, and 10000 respectively. The capacitance values used for the simulation are calculated  from the experimentally observed plateaus of $\epsilon_1$ vs $T$ plots, which yielded $C_1$ = 1.76 pF ($\epsilon_{low}$ = 20), $C_2$ = 88.4 pF ($\epsilon_{med}$ = 100), and $C_3$ = 1.326 nF ($\epsilon_{max}$ = 15000). The simulated plots for real and imaginary part of the permittivity are shown in Fig~\ref{fig_simulation} (a, b). The real part, $\epsilon_1$, has three plateaus in accordance with the  three types of relaxations. The the bulk, grain boundaries and third from the sample electrode interface\cite{Schmidt2012,Schmidt2007}. If temperature is lowered the intrinsic contribution dominates and at sufficiently low $T$, the extrinsic contributions ceases ( fig~\ref{fig_simulation} (c) ). The imaginary part $\epsilon_2$ has two peaks which is similar to the experimental data (Fig\ref{fig_epsilon}(b) and (d)). These features are also reflected in the simulated frequency dependent dielectric constant, Fig~\ref{fig_simulation}(c) and (d). \\
\begin{figure}[!t]
\centering
\includegraphics[scale=0.35]{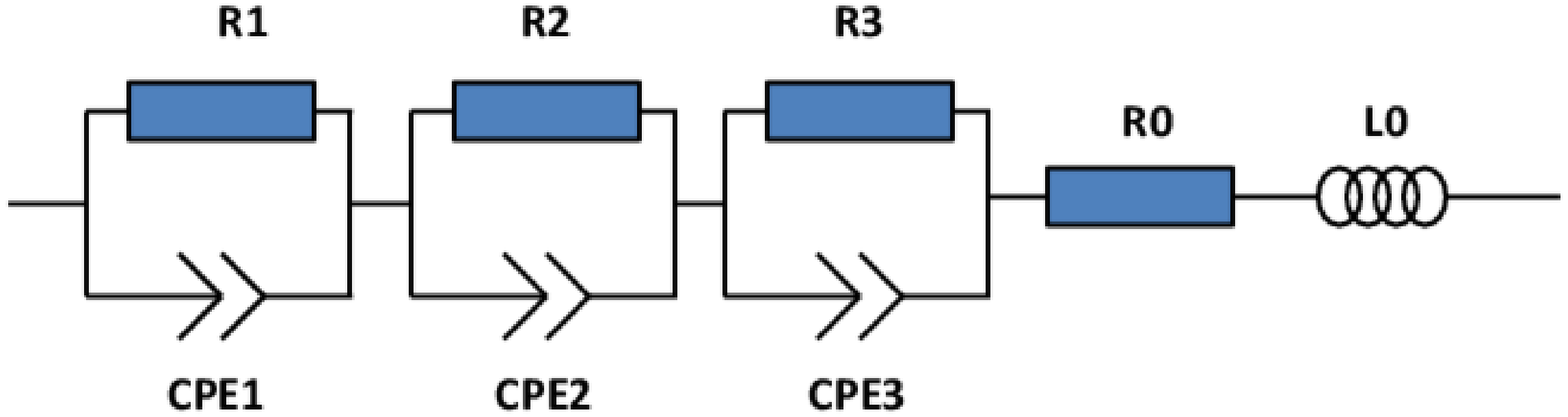}
\vspace{1cm}
\includegraphics[scale=0.35]{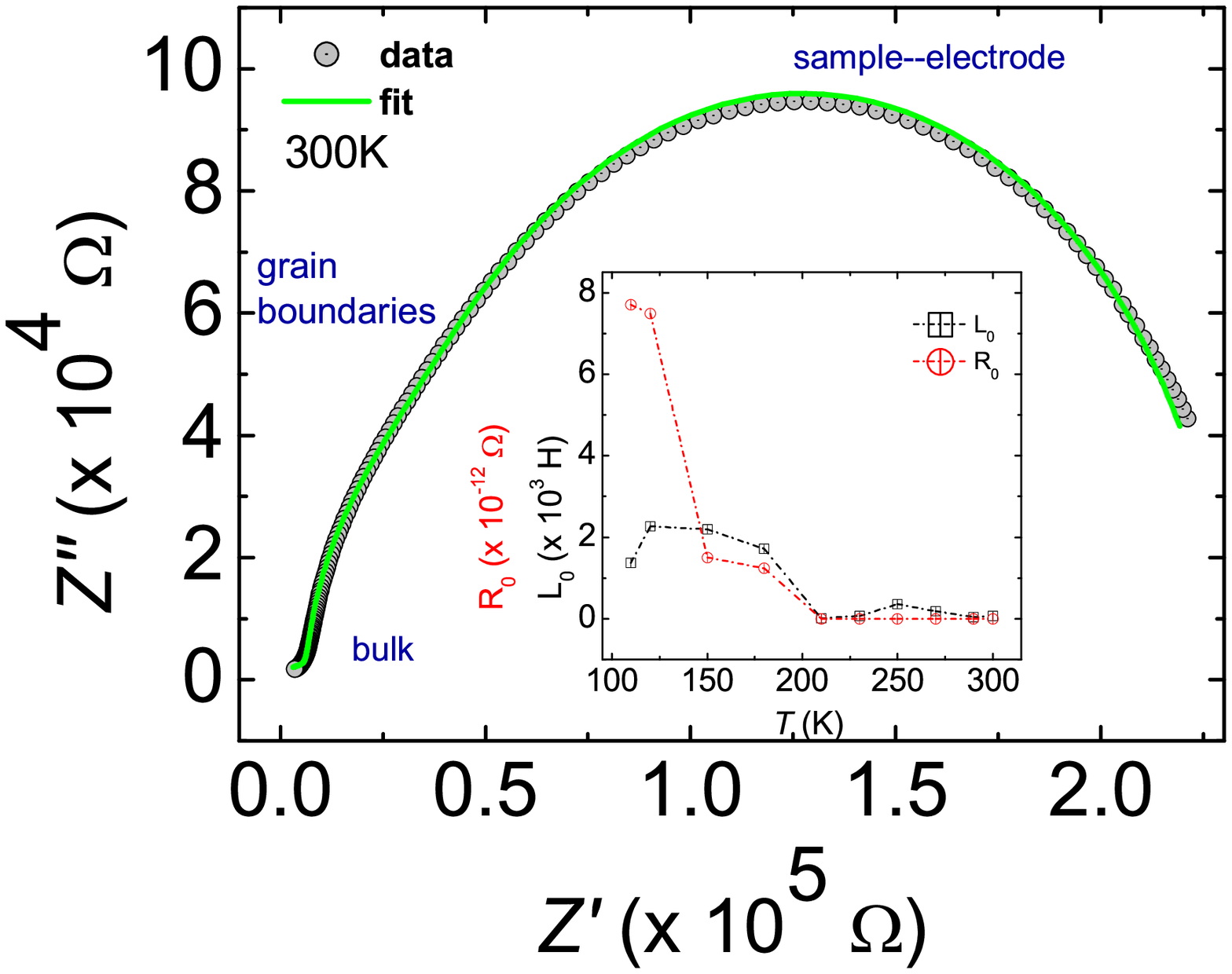}
\caption{(colour online) {\it Top:} The $R$-$CPE$ circuit used to model the complex plane of impedance, $Z'$ vs $Z"$. {\it Bottom:} The experimental $Z'$ vs $Z"$ plot at 300~K shown as circles. A fit to the observed data according to the circuit model shown above is given as solid line. The three different contributions to the total relaxation are marked.  The inset shows the temperature dependence of $R_0$ and $L_0$ (in the circuit) extracted from the fit. A clear divergence in slope is seen in both $R_0$ and $L_0$ around 200 K.}
\label{fig_RC1}
\end{figure}
\indent Impedance spectroscopy is a very powerful tool to study the multiple relaxations observed in dielectric materials, where the real ($Z'$) versus imaginary ($Z"$) part of the complex impedance for different frequencies are plotted together. Subsequent analysis using $RC$ element model can deconvolute different types of dielectric relaxations present in the material.\cite{Macdonald-Book} In the ideal case of single relaxation, the response is a semi-circle.\cite{Irvine1990} In real systems, deviation from ideal behaviour occur and in order to account for this non-Debye type behavior, the ideal capacitor is replaced with a Constant Phase Element (CPE). The complex impedance of a CPE is defined as\cite{Jonscher-Book,Schmidt2012}
\begin{equation}
Z^*_{CPE} = \frac{1}{C_{CPE}(i\omega)^n}
\end{equation}
where $C_{CPE}$ is the CPE-specific capacitance. $\omega$ is the angular frequency and $n$ is a critical exponent with typical values between 0.6 and 1 (for ideal capacitor $n$ = 1). Such CPE capacitance can be converted into real capacitance using standard procedure.\cite{Hsu2001} . Such a circuit model constructed for  the present work is illustrated in the top panel of Fig~\ref{fig_RC1}. The impedance of the sample is measured at different temperatures and $Z"$ versus $Z'$ are plotted. The room temperature impedance spectra in the complex plane is shown in Fig~\ref{fig_RC1} as black open circles. The data are fitted using three $R$-$CPE$ units in series corresponding to the bulk, the grain boundary and the sample-electrode contributions. The inductance $L_0$  of the external leads (also related to the magnetic phase) and the resistance, $R_0$, of the leads and electrodes are also taken into account in the circuit model. The curve fit to the experimental data at 300~K using this equivalent circuit is shown in Fig~\ref{fig_RC1} as solid line.
\begin{figure}[!t]
\centering
\includegraphics[scale=0.3]{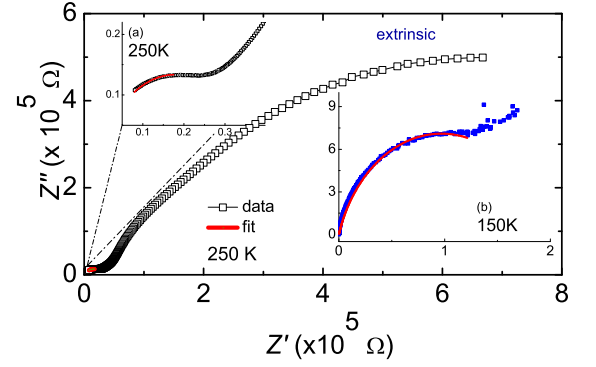}
\caption{(color online) The complex plane of $Z'$ vs $Z"$ at 250~K. The inset (a) shows an enlarged view of the region where bulk contribution (red line) dominates. The data and fit at 150~K is shown in the inset (b).}
\label{fig_RC2}
\end{figure}
The fit parameters are $R_1$ = 6467 $\pm$23\% ohm, $R_2$ = 23625 $\pm$9\% ohm, $R_3$ = 200860 $\pm$1.9\% ohm, $C_1$ = 5.33 nF $\pm$2.9\%, $C_2$ = 1.62 nF $\pm$2.6\% , $C_3$ = 2.39 pF $\pm$2.1\%, $n_1$ = 0.79 $\pm$1.1\%, $n_2$ = 0.98 $\pm$2.4\% and $n_3$ = 0.95 $\pm$1.8\%. The high frequency response originates normally from the bulk, the intermediate frequency from the grain boundaries and the low frequency impedance contribution  from the sample-electrode interface. \cite{Lunkenheimer2010} Impedance data of magnetic materials are usually modeled by including an inductive element in the equivalent circuit wherein the permiability and magnetic response can be roughly understood from the inductance.\cite{Irvin1990,Rajesh1999} A combination of $R$ and $L$ (parallel or series) are generally used for modeling the magnetic phase.\cite{Lee2008,Irvin1990} Using the circuit shown in fig~\ref{fig_RC1} we tried to extract the temperature dependence of inductance $L_0$ along with $R_0$ and is shown in the inset of bottom panel of fig~\ref{fig_RC1}. A clear anomaly is observed around 200~K which is close to the Griffith's temperature observed in this material\cite{Nair2011}. Irvin {\it et al.,} for example, have observed that the inductance value peaks at the magnetic anomaly transitions in (NiZn)Fe$_2$O$_4$.\cite{Irvin1990} \\ 
\indent The intrinsic bulk contribution is analyzed using  a single $R$-$CPE$ unit by considering the intermediate and low frequency region as extrinsic. The data is fitted faithfully upto 110~K. Fig~\ref{fig_RC2} shows the fit at 250~K (inset (a))and 150~K (inset (b)). At low temperature the bulk contribution dominates and the extrinsic contribution reduces. By a progressive procedure of fitting, the intrinsic response to the dielectric relaxation was extracted. The resistance ($R$) and the capacitance ($C$) estimated from the fit are plotted against temperature, Fig~\ref{fig_R}. A slope change is visible in the temperature region around 200~K which coincides with the experimentally observed Griffith's temperature in this compound.\cite{Nair2011} The spin-lattice coupling also shows a marked change at this point. Although one cannot claim this to be magneto-electric coupling, it certainly indicate a connection between the various degrees of freedom.
For ideal Debye relaxation of non-interacting dipoles, the plot of $\epsilon_1 (T)$ versus $\epsilon_2 (T)$ follows Cole-Cole behaviour displaying a perfect semi-circular arc \cite{Cole1941}. Like in many complex oxides and those with defects, deviations from ideal Cole-Cole plot are observed here; these may be accounted by using the modified-Debye equation, 
\begin{equation}
\epsilon^{*} = \epsilon_{1} + i \epsilon_{2} = \epsilon_\infty + \frac{(\epsilon_0 - \epsilon_\infty)}{[1 + (i\omega \tau)^{(1-\alpha)}]}
\label{Debye_eqn}
\end{equation}
where $\epsilon_0$ and $\epsilon_\infty$ are static and high frequency dielectric constants, respectively, $\omega$ is the angular frequency, $\tau$ is the mean 
relaxation time and $\alpha$ is a parameter which represents the distribution of relaxation times (for ideal Debye relaxation, $\alpha$ is zero). Equation~(\ref{Debye_eqn}) can be separated into the real and imaginary part of dielectric permittivity as,
\begin{eqnarray}
\epsilon_1 = \epsilon_\infty + (\Delta\epsilon/2) [1 - \frac{\mathrm{sinh}(\beta z)}{\mathrm{cosh}(\beta z) + \mathrm{cos}(\beta \pi/2)}]
\label{Debye_eqn1}
\\
\mathrm{and},\:\:\: \epsilon_2 = \frac{(\Delta\epsilon/2)\mathrm{sin}(\beta \pi/2)}{\mathrm{cosh}(\beta z) + \mathrm{cos}(\beta \pi/2)}
\label{Debye_eqn2}
\end{eqnarray}
where $\Delta \epsilon$ = ($\epsilon_0$ -  $\epsilon_\infty$), $z$ = ln($\omega\tau$) and $\beta$ = (1 - $\alpha$).
\begin{figure}[!t]
\centering
\includegraphics[scale=0.30]{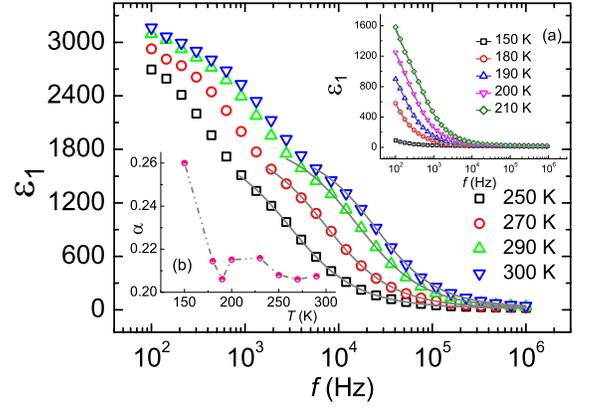}
\caption{(colour online) Frequency dependence of real part of dielectric constant, $\epsilon_1 (T)$, at different temperatures above $T^* \approx$ 200~K. The inset (a) shows $\epsilon_1 (T)$ below 250~K. The solid lines are fits using modified-Debye equation, Eqn~\ref{Debye_eqn1}. The behaviour of dielectric function is clearly different for temperatures above and below $T^*$. The inset (b) shows the variation of $\alpha$ with temperature across $T^*$.}
\label{fig_Debye}
\end{figure}
{The frequency dependence of $\epsilon_1 (T)$ for Tb$_2$NiMnO$_6$ at few select temperatures is depicted in Fig~\ref{fig_Debye} along with the curve fits using modified-Debye equation (Eqn~\ref{Debye_eqn1}). Since bulk contribution will reflect in the high frequency region, fitting was performed in the high frequency region. However, the modified-Debye model was inadequate to describe the behaviour of $\epsilon_1$ below 120~K (fig not shown}) and attempts to fit Eqn~(\ref{Debye_eqn1}) were not successful. The value extracted for $\alpha$ from the fit falls between 0.27 -- 0.18 which is higher than the value reported for La$_2$NiMnO$_6$ \cite{lin_ssc_149_784_2009}. Inset of fig~\ref{fig_Debye} (b)shows the temperature variation of $\alpha$ values extracted from the fit. An anomaly in the behavior  is observed in the vicinity of $\sim$ 200 K. We denote the temperature at which the deviation  occurs as $T ^*$, which is close to the temperature where an anomaly in the FWHM of Raman modes was observed in Tb$_2$NiMnO$_6$ (180~K) \cite{Nair2011}. This coincidence gives credence to the correlation between dielectric, Raman and magnetic anomalies in this material.\\
\indent The frequency dependence of ac conductivity $\sigma$ of Tb$_2$NiMnO$_6$ is shown in Fig~\ref{fig_accond} (a) and (b). This corresponds to two types of conduction behaviour at $T < T^*$ and $T>T^*$. Below 200~K, conductivity decreases with increase in temperature reflecting an insulating behaviour. A change of slope in ac conductivity occurs at the temperature where dielectric relaxation deviates from Debye-like behaviour. According to universal dielectric response (UDR), \cite{Jonscher-Book} the relation between conductivity $\sigma' (f)$ and dielectric constant $\epsilon_1$ is given by,
\begin{equation}
\sigma' (f) = \sigma_{dc} + \sigma_0 f^{s}
\label{udr1}
\end{equation}
and,
\begin{equation}
\epsilon_1 = \mathrm{tan}(s \pi/2) \sigma_0 f^{s-1}/ \epsilon_0
\label{udr2}
\end{equation}
where, $f$ is the experimental frequency, $\sigma_0$ and $s$ are temperature dependent constants. A step wise increase in the background of loss factor reveals the contribution from dc conductivity (fig~\ref{fig_epsilon} (b) and (d)). Equation~(\ref{udr2}) can be written as,
\begin{equation}
f \epsilon_1 = A(T) f^{s}
\label{udr3}
\end{equation}
where, $A(T)$ = $\mathrm{tan}(\frac{s\pi}{2})$ $(\frac{\sigma_0}{\epsilon_0})$,
a plot of log($f\epsilon_1$) vs log($f$) results in a straight line with slope equal to $s$. This is presented in Fig~\ref{fig_accond} (c) where a clear straight line is observed at low temperatures.
However, it deviates from linearity as temperature increases above $T^* \approx$ 200~K, when charge carriers contribute to polarization.  In order to find the dc contribution, the low frequency ac conductivity was extrapolated to zero frequency. The dc conductivity of Tb$_2$NiMnO$_6$ relates to semiconducting like behaviour above 200~K but deviates from thermally activated behaviour. The variation of dc conductivity, $\sigma_{DC}$, with temperature is shown in the inset of Fig~\ref{fig_accond} (c) where the change in slope is clearly visible close to 200~K. Again, the characteristic temperature $T^* \sim$ 200~K compares well with the the reported anomaly in the FWHM of Raman modes \cite{Nair2011}. Figure~\ref{fig_accond} (d) shows the electrical resistivity of Tb$_2$NiMnO$_6$  plotted in log scale against 1000/T. The resistivity data fits reasonably well to the Arrhenius law up to $\sim$ 200~K. This supports the dc conductivity data derived from the ac conductivity and yields activation energy, $E_a$ = 0.192(1)~eV. \\
\indent In our previous work concerning magnetic properties of Tb$_2$NiMnO$_6$, an inhomogeneous magnetic state resembling Griffiths phase was observed along with clear indications of spin-lattice coupling through Raman studies.\cite{Nair2011} We were able to extract a characteristic temperature $T^*\approx$ 200~K where magnetic and Raman anomalies coincided. The present dielectric response study validated by the dielectric relaxation and the ensuing impedence and  conductivity behavior reflect the same  characteristic temperature $T^*$  in this compound. Diffuse dielectric behaviour has been observed in nanoparticle samples of the double perovskite La$_2$NiMnO$_6$ where disorder leads to local polar nanoregions\cite{masud_jpcm_24_295902_2012observation} and also in La$_2$CoMnO$_6$ ceramics\cite{lin_jacs_94_782_2011dielectric} where the charge order of Co$^{2+}$ and Mn$^{4+}$ was the origin of relaxation. It could be assumed, in th epresent case, that clusters that posses local polarization (polar nanoregions) are formed around Ni$^{2+}$ or Mn$^{4+}$ due to the cationic antisite disorder and bring about dielectric relaxation through their mutual interaction. This is also supported by the fact that the frequency dependence of the peak temperatute in dielectric data is not explained by thermally activated behaviour. Similar to the magnetic Griffiths phase where the nonanalyticity of magnetization extends above the transition temperature (in the present case, $T_c \approx$ 110~K), the dielectric relaxation also identifies the characteristic temperature above $T_c$. In this context, it would be intersting to perform experimental investigations using local probes like electron paramagnetic resonance which can give clear signatures about the Griffiths phase\cite{deisenhofer_prl_95_257202_2005observation}. Combining this with the fact that the Griffiths phase can indeed make itself manifested in disordered dielectrics\cite{stephanovich_epjb_18_17_2000griffiths} makes this double perovskite an interesting candidate to carry out microscopic measurements, for example using neutrons, to understand the magnetic and dielectric properties in more detail. Impedance spectroscopy and modified-Debye model is used to analysis the dielectric response and identified a characteristic temperature that separates two regions of dielectric relaxation. This allows us to postulate a close correlation between magnetic, electronic and dielectric properties in this compound.
\\
\section{Acknowledgements}
SE wishes to acknowledge Department of Science and Technology, India for financial
support. HN wishes to acknowledge Aditya Wagh for his help and support with
dielectric measurements.\\ \\
%

\begin{thebibliography}{24}%
\makeatletter
\providecommand \@ifxundefined [1]{%
 \@ifx{#1\undefined}
}%
\providecommand \@ifnum [1]{%
 \ifnum #1\expandafter \@firstoftwo
 \else \expandafter \@secondoftwo
 \fi
}%
\providecommand \@ifx [1]{%
 \ifx #1\expandafter \@firstoftwo
 \else \expandafter \@secondoftwo
 \fi
}%
\providecommand \natexlab [1]{#1}%
\providecommand \enquote  [1]{``#1''}%
\providecommand \bibnamefont  [1]{#1}%
\providecommand \bibfnamefont [1]{#1}%
\providecommand \citenamefont [1]{#1}%
\providecommand \href@noop [0]{\@secondoftwo}%
\providecommand \href [0]{\begingroup \@sanitize@url \@href}%
\providecommand \@href[1]{\@@startlink{#1}\@@href}%
\providecommand \@@href[1]{\endgroup#1\@@endlink}%
\providecommand \@sanitize@url [0]{\catcode `\\12\catcode `\$12\catcode
  `\&12\catcode `\#12\catcode `\^12\catcode `\_12\catcode `\%12\relax}%
\providecommand \@@startlink[1]{}%
\providecommand \@@endlink[0]{}%
\providecommand \url  [0]{\begingroup\@sanitize@url \@url }%
\providecommand \@url [1]{\endgroup\@href {#1}{\urlprefix }}%
\providecommand \urlprefix  [0]{URL }%
\providecommand \Eprint [0]{\href }%
\providecommand \doibase [0]{http://dx.doi.org/}%
\providecommand \selectlanguage [0]{\@gobble}%
\providecommand \bibinfo  [0]{\@secondoftwo}%
\providecommand \bibfield  [0]{\@secondoftwo}%
\providecommand \translation [1]{[#1]}%
\providecommand \BibitemOpen [0]{}%
\providecommand \bibitemStop [0]{}%
\providecommand \bibitemNoStop [0]{.\EOS\space}%
\providecommand \EOS [0]{\spacefactor3000\relax}%
\providecommand \BibitemShut  [1]{\csname bibitem#1\endcsname}%
\let\auto@bib@innerbib\@empty
\bibitem [{\citenamefont {Dass}, \citenamefont {Yan},\ and\ \citenamefont
  {Goodenough}(2003)}]{Dass2003}%
  \BibitemOpen
  \bibfield  {author} {\bibinfo {author} {\bibfnamefont {R.~I.}\ \bibnamefont
  {Dass}}, \bibinfo {author} {\bibfnamefont {J.~Q.}\ \bibnamefont {Yan}}, \
  and\ \bibinfo {author} {\bibfnamefont {J.~B.}\ \bibnamefont {Goodenough}},\
  }\href@noop {} {\bibfield  {journal} {\bibinfo  {journal} {Phy. Rev. B}\
  }\textbf {\bibinfo {volume} {68}},\ \bibinfo {pages} {064415} (\bibinfo
  {year} {2003})}\BibitemShut {NoStop}%
\bibitem [{\citenamefont {Rogado}\ \emph {et~al.}(2005)\citenamefont {Rogado},
  \citenamefont {Li}, \citenamefont {Sleight},\ and\ \citenamefont
  {Subramanian}}]{Rogado2005}%
  \BibitemOpen
  \bibfield  {author} {\bibinfo {author} {\bibfnamefont {N.~S.}\ \bibnamefont
  {Rogado}}, \bibinfo {author} {\bibfnamefont {J.}~\bibnamefont {Li}}, \bibinfo
  {author} {\bibfnamefont {A.~W.}\ \bibnamefont {Sleight}}, \ and\ \bibinfo
  {author} {\bibfnamefont {M.~A.}\ \bibnamefont {Subramanian}},\ }\href@noop {}
  {\bibfield  {journal} {\bibinfo  {journal} {Adv. Mater.}\ }\textbf {\bibinfo
  {volume} {17}},\ \bibinfo {pages} {2225} (\bibinfo {year}
  {2005})}\BibitemShut {NoStop}%
\bibitem [{\citenamefont {Singh}\ \emph {et~al.}(2007)\citenamefont {Singh},
  \citenamefont {Grygiel}, \citenamefont {Sheets}, \citenamefont {Boullay},
  \citenamefont {Hervieu}, \citenamefont {Prellier}, \citenamefont {Mercey},
  \citenamefont {Simon},\ and\ \citenamefont {Raveau}}]{Singh2007}%
  \BibitemOpen
  \bibfield  {author} {\bibinfo {author} {\bibfnamefont {M.~P.}\ \bibnamefont
  {Singh}}, \bibinfo {author} {\bibfnamefont {C.}~\bibnamefont {Grygiel}},
  \bibinfo {author} {\bibfnamefont {W.~C.}\ \bibnamefont {Sheets}}, \bibinfo
  {author} {\bibfnamefont {P.}~\bibnamefont {Boullay}}, \bibinfo {author}
  {\bibfnamefont {M.}~\bibnamefont {Hervieu}}, \bibinfo {author} {\bibfnamefont
  {W.}~\bibnamefont {Prellier}}, \bibinfo {author} {\bibfnamefont
  {B.}~\bibnamefont {Mercey}}, \bibinfo {author} {\bibfnamefont
  {C.}~\bibnamefont {Simon}}, \ and\ \bibinfo {author} {\bibfnamefont
  {B.}~\bibnamefont {Raveau}},\ }\href@noop {} {\bibfield  {journal} {\bibinfo
  {journal} {Appl. Phys. Lett.}\ }\textbf {\bibinfo {volume} {91}},\ \bibinfo
  {pages} {012503} (\bibinfo {year} {2007})}\BibitemShut {NoStop}%
\bibitem [{\citenamefont {Kumar}, \citenamefont {Giovannetti},\ and\
  \citenamefont {van~den Brink}(2010)}]{Kumar2010}%
  \BibitemOpen
  \bibfield  {author} {\bibinfo {author} {\bibfnamefont {S.}~\bibnamefont
  {Kumar}}, \bibinfo {author} {\bibfnamefont {G.}~\bibnamefont {Giovannetti}},
  \ and\ \bibinfo {author} {\bibfnamefont {S.}~\bibnamefont {van~den Brink},
  \bibfnamefont {J.and~Picozzi}},\ }\href@noop {} {\bibfield  {journal}
  {\bibinfo  {journal} {Phys. Rev. B}\ }\textbf {\bibinfo {volume} {82}},\
  \bibinfo {pages} {134429} (\bibinfo {year} {2010})}\BibitemShut {NoStop}%
\bibitem [{\citenamefont {Y\'{a}\~{n}ez Vilar}\ \emph
  {et~al.}(2011)\citenamefont {Y\'{a}\~{n}ez Vilar}, \citenamefont {Mun},
  \citenamefont {Zapf}, \citenamefont {Ueland}, \citenamefont {Gardner},
  \citenamefont {Thompson}, \citenamefont {Singleton}, \citenamefont
  {S\'{a}nchez-And\'{u}jar}, \citenamefont {Mira}, \citenamefont {Biskup},
  \citenamefont {Se\~{n}ar\'{\i}s Rodr\'{\i}guez},\ and\ \citenamefont
  {Batista}}]{Yanez-Vilar2011}%
  \BibitemOpen
  \bibfield  {author} {\bibinfo {author} {\bibfnamefont {S.}~\bibnamefont
  {Y\'{a}\~{n}ez Vilar}}, \bibinfo {author} {\bibfnamefont {E.~D.}\
  \bibnamefont {Mun}}, \bibinfo {author} {\bibfnamefont {V.~S.}\ \bibnamefont
  {Zapf}}, \bibinfo {author} {\bibfnamefont {B.~G.}\ \bibnamefont {Ueland}},
  \bibinfo {author} {\bibfnamefont {J.~S.}\ \bibnamefont {Gardner}}, \bibinfo
  {author} {\bibfnamefont {J.~D.}\ \bibnamefont {Thompson}}, \bibinfo {author}
  {\bibfnamefont {J.}~\bibnamefont {Singleton}}, \bibinfo {author}
  {\bibfnamefont {M.}~\bibnamefont {S\'{a}nchez-And\'{u}jar}}, \bibinfo
  {author} {\bibfnamefont {J.}~\bibnamefont {Mira}}, \bibinfo {author}
  {\bibfnamefont {N.}~\bibnamefont {Biskup}}, \bibinfo {author} {\bibfnamefont
  {M.~A.}\ \bibnamefont {Se\~{n}ar\'{\i}s Rodr\'{\i}guez}}, \ and\ \bibinfo
  {author} {\bibfnamefont {C.~D.}\ \bibnamefont {Batista}},\ }\href@noop {}
  {\bibfield  {journal} {\bibinfo  {journal} {Phys. Rev. B}\ }\textbf {\bibinfo
  {volume} {84}},\ \bibinfo {pages} {134427} (\bibinfo {year}
  {2011})}\BibitemShut {NoStop}%
\bibitem [{\citenamefont {Padhan}\ \emph {et~al.}(2008)\citenamefont {Padhan},
  \citenamefont {Guo}, \citenamefont {LeClair},\ and\ \citenamefont
  {Gupta}}]{Padhan2008}%
  \BibitemOpen
  \bibfield  {author} {\bibinfo {author} {\bibfnamefont {P.}~\bibnamefont
  {Padhan}}, \bibinfo {author} {\bibfnamefont {H.~Z.}\ \bibnamefont {Guo}},
  \bibinfo {author} {\bibfnamefont {P.}~\bibnamefont {LeClair}}, \ and\
  \bibinfo {author} {\bibfnamefont {A.}~\bibnamefont {Gupta}},\ }\href@noop {}
  {\bibfield  {journal} {\bibinfo  {journal} {App. Phys. Lett.}\ }\textbf
  {\bibinfo {volume} {92}},\ \bibinfo {pages} {022909} (\bibinfo {year}
  {2008})}\BibitemShut {NoStop}%
\bibitem [{\citenamefont {Lin}, \citenamefont {Chen},\ and\ \citenamefont
  {Liu}(2009)}]{lin_ssc_149_784_2009}%
  \BibitemOpen
  \bibfield  {author} {\bibinfo {author} {\bibfnamefont {Y.~Q.}\ \bibnamefont
  {Lin}}, \bibinfo {author} {\bibfnamefont {X.~M.}\ \bibnamefont {Chen}}, \
  and\ \bibinfo {author} {\bibfnamefont {X.~Q.}\ \bibnamefont {Liu}},\
  }\href@noop {} {\bibfield  {journal} {\bibinfo  {journal} {Solid State
  Commun.}\ }\textbf {\bibinfo {volume} {149}},\ \bibinfo {pages} {784}
  (\bibinfo {year} {2009})}\BibitemShut {NoStop}%
\bibitem [{\citenamefont {Nair}\ \emph {et~al.}(2011)\citenamefont {Nair},
  \citenamefont {Swain}, \citenamefont {Hariharan}, \citenamefont {Adiga},
  \citenamefont {Narayana},\ and\ \citenamefont {Elizabeth}}]{Nair2011}%
  \BibitemOpen
  \bibfield  {author} {\bibinfo {author} {\bibfnamefont {H.~S.}\ \bibnamefont
  {Nair}}, \bibinfo {author} {\bibfnamefont {D.}~\bibnamefont {Swain}},
  \bibinfo {author} {\bibfnamefont {N.}~\bibnamefont {Hariharan}}, \bibinfo
  {author} {\bibfnamefont {S.}~\bibnamefont {Adiga}}, \bibinfo {author}
  {\bibfnamefont {C.}~\bibnamefont {Narayana}}, \ and\ \bibinfo {author}
  {\bibfnamefont {S.}~\bibnamefont {Elizabeth}},\ }\href@noop {} {\bibfield
  {journal} {\bibinfo  {journal} {J. Appl. Phys.}\ }\textbf {\bibinfo {volume}
  {110}},\ \bibinfo {pages} {123919} (\bibinfo {year} {2011})}\BibitemShut
  {NoStop}%
\bibitem [{\citenamefont {Schmidt}\ \emph {et~al.}(2012)\citenamefont
  {Schmidt}, \citenamefont {Ventura}, \citenamefont {Langenberg}, \citenamefont
  {Nemes}, \citenamefont {Munuera}, \citenamefont {Varela}, \citenamefont
  {Garcia-Hernandez}, \citenamefont {Leon}, \citenamefont {Santamaria},\ and\
  \citenamefont {Kelvin}}]{Schmidt2012}%
  \BibitemOpen
  \bibfield  {author} {\bibinfo {author} {\bibfnamefont {R.}~\bibnamefont
  {Schmidt}}, \bibinfo {author} {\bibfnamefont {J.}~\bibnamefont {Ventura}},
  \bibinfo {author} {\bibfnamefont {E.}~\bibnamefont {Langenberg}}, \bibinfo
  {author} {\bibfnamefont {N.~M.}\ \bibnamefont {Nemes}}, \bibinfo {author}
  {\bibfnamefont {C.}~\bibnamefont {Munuera}}, \bibinfo {author} {\bibfnamefont
  {M.}~\bibnamefont {Varela}}, \bibinfo {author} {\bibfnamefont
  {M.}~\bibnamefont {Garcia-Hernandez}}, \bibinfo {author} {\bibfnamefont
  {C.}~\bibnamefont {Leon}}, \bibinfo {author} {\bibfnamefont {J.}~\bibnamefont
  {Santamaria}}, \ and\ \bibinfo {author} {\bibfnamefont {T.}~\bibnamefont
  {Kelvin}},\ }\href@noop {} {\bibfield  {journal} {\bibinfo  {journal} {Phy.
  Rev. B}\ }\textbf {\bibinfo {volume} {86}},\ \bibinfo {pages} {035113}
  (\bibinfo {year} {2012})}\BibitemShut {NoStop}%
\bibitem [{\citenamefont {Baron-Gonzalez}\ \emph {et~al.}(2011)\citenamefont
  {Baron-Gonzalez}, \citenamefont {Frontera}, \citenamefont
  {Grac\'{i}a-Mu\~{n}oz},\ and\ \citenamefont
  {Rivas-Murias}}]{Baron-Gonzalez2011}%
  \BibitemOpen
  \bibfield  {author} {\bibinfo {author} {\bibfnamefont {A.~J.}\ \bibnamefont
  {Baron-Gonzalez}}, \bibinfo {author} {\bibfnamefont {C.}~\bibnamefont
  {Frontera}}, \bibinfo {author} {\bibfnamefont {J.~L.}\ \bibnamefont
  {Grac\'{i}a-Mu\~{n}oz}}, \ and\ \bibinfo {author} {\bibfnamefont
  {J.}~\bibnamefont {Rivas-Murias}, \bibfnamefont {Band~Blasco}},\ }\href@noop
  {} {\bibfield  {journal} {\bibinfo  {journal} {J. Phys.: Condens. Matter}\
  }\textbf {\bibinfo {volume} {23}},\ \bibinfo {pages} {496003} (\bibinfo
  {year} {2011})}\BibitemShut {NoStop}%
\bibitem [{\citenamefont {Schmidt}\ \emph {et~al.}(2007)\citenamefont
  {Schmidt}, \citenamefont {Eerenstein}, \citenamefont {Winiecki},
  \citenamefont {Morrison},\ and\ \citenamefont {Midgley}}]{Schmidt2007}%
  \BibitemOpen
  \bibfield  {author} {\bibinfo {author} {\bibfnamefont {R.}~\bibnamefont
  {Schmidt}}, \bibinfo {author} {\bibfnamefont {W.}~\bibnamefont {Eerenstein}},
  \bibinfo {author} {\bibfnamefont {T.}~\bibnamefont {Winiecki}}, \bibinfo
  {author} {\bibfnamefont {F.~D.}\ \bibnamefont {Morrison}}, \ and\ \bibinfo
  {author} {\bibfnamefont {P.~A.}\ \bibnamefont {Midgley}},\ }\href@noop {}
  {\bibfield  {journal} {\bibinfo  {journal} {Phys. Rev. B}\ }\textbf {\bibinfo
  {volume} {75}},\ \bibinfo {pages} {245111} (\bibinfo {year}
  {2007})}\BibitemShut {NoStop}%
\bibitem [{\citenamefont {Lunkenheimer}\ \emph {et~al.}(2010)\citenamefont
  {Lunkenheimer}, \citenamefont {Krohns}, \citenamefont {Riegg}, \citenamefont
  {Ebbinghaus}, \citenamefont {Reller},\ and\ \citenamefont
  {Loidl}}]{Lunkenheimer2010}%
  \BibitemOpen
  \bibfield  {author} {\bibinfo {author} {\bibfnamefont {P.}~\bibnamefont
  {Lunkenheimer}}, \bibinfo {author} {\bibfnamefont {S.}~\bibnamefont
  {Krohns}}, \bibinfo {author} {\bibfnamefont {S.}~\bibnamefont {Riegg}},
  \bibinfo {author} {\bibfnamefont {S.~G.}\ \bibnamefont {Ebbinghaus}},
  \bibinfo {author} {\bibfnamefont {A.}~\bibnamefont {Reller}}, \ and\ \bibinfo
  {author} {\bibfnamefont {A.}~\bibnamefont {Loidl}},\ }\href@noop {}
  {\bibfield  {journal} {\bibinfo  {journal} {Eur. Phys. J. Special Topics}\
  }\textbf {\bibinfo {volume} {89}},\ \bibinfo {pages} {61} (\bibinfo {year}
  {2010})}\BibitemShut {NoStop}%
\bibitem [{\citenamefont {Barsukov}\ and\ \citenamefont
  {Macdonald}(2005)}]{Macdonald-Book}%
  \BibitemOpen
  \bibfield  {author} {\bibinfo {author} {\bibfnamefont {E.}~\bibnamefont
  {Barsukov}}\ and\ \bibinfo {author} {\bibfnamefont {J.}~\bibnamefont
  {Macdonald}},\ }\href@noop {} {\emph {\bibinfo {title} {Impedance
  Spectroscopy; Theory, Ecperiment and Applications}}}\ (\bibinfo  {publisher}
  {Wiley, Hoboken, New Jersey},\ \bibinfo {year} {2005})\BibitemShut {NoStop}%
\bibitem [{\citenamefont {Irvine}, \citenamefont {Sinclair},\ and\
  \citenamefont {West}(1990)}]{Irvine1990}%
  \BibitemOpen
  \bibfield  {author} {\bibinfo {author} {\bibfnamefont {J.~T.~S.}\
  \bibnamefont {Irvine}}, \bibinfo {author} {\bibfnamefont {D.~C.}\
  \bibnamefont {Sinclair}}, \ and\ \bibinfo {author} {\bibfnamefont {A.~R.}\
  \bibnamefont {West}},\ }\href@noop {} {\bibfield  {journal} {\bibinfo
  {journal} {Adv. Mater.}\ }\textbf {\bibinfo {volume} {2}},\ \bibinfo {pages}
  {132} (\bibinfo {year} {1990})}\BibitemShut {NoStop}%
\bibitem [{\citenamefont {Jonscher}(1983)}]{Jonscher-Book}%
  \BibitemOpen
  \bibfield  {author} {\bibinfo {author} {\bibfnamefont {A.~K.}\ \bibnamefont
  {Jonscher}},\ }\href@noop {} {\emph {\bibinfo {title} {Dielectric Relaxation
  in Solids}}}\ (\bibinfo  {publisher} {Chelsea Dielectric Press, London},\
  \bibinfo {year} {1983})\BibitemShut {NoStop}%
\bibitem [{\citenamefont {Hsu}\ and\ \citenamefont {Mansfeld}(2011)}]{Hsu2001}%
  \BibitemOpen
  \bibfield  {author} {\bibinfo {author} {\bibfnamefont {C.~H.}\ \bibnamefont
  {Hsu}}\ and\ \bibinfo {author} {\bibfnamefont {F.}~\bibnamefont {Mansfeld}},\
  }\href@noop {} {\bibfield  {journal} {\bibinfo  {journal} {Corrosios}\
  }\textbf {\bibinfo {volume} {57}},\ \bibinfo {pages} {747} (\bibinfo {year}
  {2011})}\BibitemShut {NoStop}%
\bibitem [{\citenamefont {Irvine}\ and\ \citenamefont
  {West}(1990)}]{Irvin1990}%
  \BibitemOpen
  \bibfield  {author} {\bibinfo {author} {\bibfnamefont {J.~T.~S.}\
  \bibnamefont {Irvine}}\ and\ \bibinfo {author} {\bibfnamefont {A.~R.}\
  \bibnamefont {West}},\ }\href@noop {} {\bibfield  {journal} {\bibinfo
  {journal} {Solid State Ionics}\ }\textbf {\bibinfo {volume} {40/41}},\
  \bibinfo {pages} {220} (\bibinfo {year} {1990})}\BibitemShut {NoStop}%
\bibitem [{\citenamefont {Katare}\ \emph {et~al.}(1999)\citenamefont {Katare},
  \citenamefont {Pandey}, \citenamefont {Dwivedi}, \citenamefont {Prakash},\
  and\ \citenamefont {Kumar}}]{Rajesh1999}%
  \BibitemOpen
  \bibfield  {author} {\bibinfo {author} {\bibfnamefont {R.~K.}\ \bibnamefont
  {Katare}}, \bibinfo {author} {\bibfnamefont {L.}~\bibnamefont {Pandey}},
  \bibinfo {author} {\bibfnamefont {R.~K.}\ \bibnamefont {Dwivedi}}, \bibinfo
  {author} {\bibfnamefont {O.}~\bibnamefont {Prakash}}, \ and\ \bibinfo
  {author} {\bibfnamefont {D.}~\bibnamefont {Kumar}},\ }\href@noop {}
  {\bibfield  {journal} {\bibinfo  {journal} {Indian Journal Engineering \&
  Material Science}\ }\textbf {\bibinfo {volume} {6}},\ \bibinfo {pages} {34}
  (\bibinfo {year} {1999})}\BibitemShut {NoStop}%
\bibitem [{\citenamefont {Lee}\ \emph {et~al.}(2008)\citenamefont {Lee},
  \citenamefont {Chou}, \citenamefont {Hsu}, \citenamefont {Lin},\ and\
  \citenamefont {Fu}}]{Lee2008}%
  \BibitemOpen
  \bibfield  {author} {\bibinfo {author} {\bibfnamefont {J.~H.}\ \bibnamefont
  {Lee}}, \bibinfo {author} {\bibfnamefont {H.}~\bibnamefont {Chou}}, \bibinfo
  {author} {\bibfnamefont {H.~S.}\ \bibnamefont {Hsu}}, \bibinfo {author}
  {\bibfnamefont {C.~P.}\ \bibnamefont {Lin}}, \ and\ \bibinfo {author}
  {\bibfnamefont {C.~M.}\ \bibnamefont {Fu}},\ }\href@noop {} {\bibfield
  {journal} {\bibinfo  {journal} {IEEE TRANSACTIONS ON MAGNETICS}\ }\textbf
  {\bibinfo {volume} {44}},\ \bibinfo {pages} {11} (\bibinfo {year}
  {2008})}\BibitemShut {NoStop}%
\bibitem [{\citenamefont {Cole}\ and\ \citenamefont {Cole}(1941)}]{Cole1941}%
  \BibitemOpen
  \bibfield  {author} {\bibinfo {author} {\bibfnamefont {K.~S.}\ \bibnamefont
  {Cole}}\ and\ \bibinfo {author} {\bibfnamefont {R.~H.}\ \bibnamefont
  {Cole}},\ }\href@noop {} {\bibfield  {journal} {\bibinfo  {journal} {J. Chem.
  Phys.}\ }\textbf {\bibinfo {volume} {9}},\ \bibinfo {pages} {341} (\bibinfo
  {year} {1941})}\BibitemShut {NoStop}%
\bibitem [{\citenamefont {Masud}\ \emph {et~al.}(2012)\citenamefont {Masud},
  \citenamefont {Ghosh}, \citenamefont {Sannigrahi},\ and\ \citenamefont
  {Chaudhuri}}]{masud_jpcm_24_295902_2012observation}%
  \BibitemOpen
  \bibfield  {author} {\bibinfo {author} {\bibfnamefont {M.~G.}\ \bibnamefont
  {Masud}}, \bibinfo {author} {\bibfnamefont {A.}~\bibnamefont {Ghosh}},
  \bibinfo {author} {\bibfnamefont {J.}~\bibnamefont {Sannigrahi}}, \ and\
  \bibinfo {author} {\bibfnamefont {B.~K.}\ \bibnamefont {Chaudhuri}},\
  }\href@noop {} {\bibfield  {journal} {\bibinfo  {journal} {J. Phys.: Condens.
  Matter}\ }\textbf {\bibinfo {volume} {24}},\ \bibinfo {pages} {295902}
  (\bibinfo {year} {2012})}\BibitemShut {NoStop}%
\bibitem [{\citenamefont {Lin}\ and\ \citenamefont
  {Chen}(2011)}]{lin_jacs_94_782_2011dielectric}%
  \BibitemOpen
  \bibfield  {author} {\bibinfo {author} {\bibfnamefont {Y.~Q.}\ \bibnamefont
  {Lin}}\ and\ \bibinfo {author} {\bibfnamefont {X.~M.}\ \bibnamefont {Chen}},\
  }\href@noop {} {\bibfield  {journal} {\bibinfo  {journal} {J. Am. Ceram.
  Soc.}\ }\textbf {\bibinfo {volume} {94}},\ \bibinfo {pages} {782} (\bibinfo
  {year} {2011})}\BibitemShut {NoStop}%
\bibitem [{\citenamefont {Deisenhofer}\ \emph {et~al.}(2005)\citenamefont
  {Deisenhofer}, \citenamefont {Braak}, \citenamefont {von Nidda},
  \citenamefont {Hemberger}, \citenamefont {Eremina}, \citenamefont {Ivanshin},
  \citenamefont {Balbashov}, \citenamefont {Jug}, \citenamefont {Loidl},
  \citenamefont {Kimura},\ and\ \citenamefont
  {Tokura}}]{deisenhofer_prl_95_257202_2005observation}%
  \BibitemOpen
  \bibfield  {author} {\bibinfo {author} {\bibfnamefont {J.}~\bibnamefont
  {Deisenhofer}}, \bibinfo {author} {\bibfnamefont {D.}~\bibnamefont {Braak}},
  \bibinfo {author} {\bibfnamefont {H.~A.~K.}\ \bibnamefont {von Nidda}},
  \bibinfo {author} {\bibfnamefont {J.}~\bibnamefont {Hemberger}}, \bibinfo
  {author} {\bibfnamefont {R.~M.}\ \bibnamefont {Eremina}}, \bibinfo {author}
  {\bibfnamefont {V.~A.}\ \bibnamefont {Ivanshin}}, \bibinfo {author}
  {\bibfnamefont {A.~M.}\ \bibnamefont {Balbashov}}, \bibinfo {author}
  {\bibfnamefont {G.}~\bibnamefont {Jug}}, \bibinfo {author} {\bibfnamefont
  {A.}~\bibnamefont {Loidl}}, \bibinfo {author} {\bibfnamefont
  {T.}~\bibnamefont {Kimura}}, \ and\ \bibinfo {author} {\bibfnamefont
  {Y.}~\bibnamefont {Tokura}},\ }\href@noop {} {\bibfield  {journal} {\bibinfo
  {journal} {Phys. Rev. Lett.}\ }\textbf {\bibinfo {volume} {95}},\ \bibinfo
  {pages} {257202} (\bibinfo {year} {2005})}\BibitemShut {NoStop}%
\bibitem [{\citenamefont
  {Stephanovich}(2000)}]{stephanovich_epjb_18_17_2000griffiths}%
  \BibitemOpen
  \bibfield  {author} {\bibinfo {author} {\bibfnamefont {V.~A.}\ \bibnamefont
  {Stephanovich}},\ }\href@noop {} {\bibfield  {journal} {\bibinfo  {journal}
  {Eur. Phys. J. B}\ }\textbf {\bibinfo {volume} {18}},\ \bibinfo {pages} {17}
  (\bibinfo {year} {2000})}\BibitemShut {NoStop}%
\end{thebibliography}
%

\end{document}